\documentclass[
	aps, prl, reprint,
	10pt, notitlepage, a4paper,
        floats, floatfix,
	amsmath, amssymb, amsfonts, 
	superscriptaddress,
	showpacs, showkeys,
	nofootinbib,
]{revtex4-1}

\def\paper{Letter}
\renewcommand{\section}[1]{\paragraph{#1. ---}\phantomsection\addcontentsline{toc}{section}{#1}}

\usepackage{ifthen}
\newboolean{prd}
\setboolean{prd}{false}
\newboolean{arxiv}
\setboolean{arxiv}{true}

\newboolean{notprd}
\setboolean{notprd}{true}
\ifprd
\setboolean{notprd}{false}
\fi
\newboolean{notarxiv}
\setboolean{notarxiv}{true}
\ifarxiv
\setboolean{notarxiv}{false}
\fi

\usepackage{graphicx}
\usepackage[usenames]{xcolor}
\usepackage{subfigure}

\xdefinecolor{mylinkcolor}{rgb}{0,0,0.5}
\usepackage[
	bookmarksnumbered, bookmarksopen, bookmarksopenlevel=2,
	breaklinks=true,
	colorlinks=true, filecolor=mylinkcolor, citecolor=mylinkcolor,
	linkcolor=mylinkcolor, urlcolor=mylinkcolor, menucolor=mylinkcolor,
]{hyperref}

\usepackage{wasysym}

\newcommand{\be}{\begin{equation}}
\newcommand{\ee}{\end{equation}}
\newcommand{\bea}{\begin{eqnarray}}
\newcommand{\eea}{\end{eqnarray}}

\newcommand{\Q}{\hat Q}
\newcommand{\I}{\hat I}
\def\f{\tilde{f}}
\def\fM{\hat{f}}
\def\R{\hat{R}}

\DeclareMathOperator{\Order}{\mathcal{O}}

\allowdisplaybreaks

\begin{document}

\hypersetup{
	pdftitle={I-Q Relation for Rapidly Rotating Neutron Stars},
	pdfauthor={Sayan Chakrabarti, Terence Delsate, Norman Guerlebeck, Jan Steinhoff}
}

\title{I-Q relation for rapidly rotating neutron stars}

\author{Sayan Chakrabarti}
\affiliation{Department of Physics, Indian Institute of Technology Guwahati,  North Guwahati, 781039, Assam, India}
\affiliation{Centro Multidisciplinar de Astrof\'isica --- CENTRA, Departamento de F\'isica,
	Instituto Superior T\'ecnico --- IST, Universidade de Lisboa --- ULisboa,
	Avenida Rovisco Pais 1, 1049-001 Lisboa, Portugal, EU}

\author{T\'erence Delsate}
\affiliation{Centro Multidisciplinar de Astrof\'isica --- CENTRA, Departamento de F\'isica,
	Instituto Superior T\'ecnico --- IST, Universidade de Lisboa --- ULisboa,
	Avenida Rovisco Pais 1, 1049-001 Lisboa, Portugal, EU}
\affiliation{UMons, Universit\'e de Mons, Place du Parc 20, 7000 Mons, Belgium, EU}

\author{Norman G{\"u}rlebeck}
\affiliation{ZARM, University of Bremen, Am Fallturm, 28359 Bremen, Germany, EU}

\author{Jan Steinhoff}
\email[Corresponding author: ]{jan.steinhoff@ist.utl.pt}
\affiliation{Centro Multidisciplinar de Astrof\'isica --- CENTRA, Departamento de F\'isica,
	Instituto Superior T\'ecnico --- IST, Universidade de Lisboa --- ULisboa,
	Avenida Rovisco Pais 1, 1049-001 Lisboa, Portugal, EU}

\date{\today}

\begin{abstract}
We consider a universal relation between moment of inertia and quadrupole moment of
arbitrarily fast rotating neutron stars.
Recent studies suggest that this relation breaks down for fast rotation.
We find that it is still universal among various suggested equations of state for constant values of certain dimensionless parameters characterizing the magnitude of rotation.
One of these parameters includes the neutron star radius,
leading to a new universal relation expressing the radius through the mass, frequency, and spin parameter.
This can become a powerful tool for radius measurements.
\end{abstract}


\maketitle

\section{Introduction}
General relativity is the cornerstone for our understanding of
gravity, one of the fundamental forces of nature. Neutron stars (NSs) are among
the best laboratories for testing strong gravity and probing other fundamental
interactions (e.g. strong interactions). To date, pulsar observations deliver
some of the best tests of general relativity and alternative theories of
gravity. In the near future, currently planned or approved observatories such as
SKA, ATHENA, LOFT and NICER will improve these tests by orders of magnitude.
Additionally, inspiraling NS binaries are the most likely sources
for gravitational wave (GW) detectors such as the Advanced LIGO, which will start
operation very soon. To exploit all these ambitious astronomical projects to full
extent, a careful and precise modeling is required.


However, a catch of using NSs for gravity tests (see, e.g., Refs.\ \cite{Will:2006, Yunes:Siemens:2013} for reviews on gravity tests) is our
current ignorance about many aspects of their structure. While there is
increasing agreement on their very outer layers among
various groups, diverse theoretical models for their inner structure are proposed.
This is due to the quantum chromodynamical interactions of the matter in
regimes not currently accessible by earth-based experiments. For this reason, theoretical
predictions are difficult.
The issue above raises the question of whether a NS can be used for precision tests
of gravity theory at all? For instance, alternative theories of gravity
can have experimental signatures similar to finite size effects (internal structure).
This was explicitly shown in, e.g.,
Ref.\ \cite{Delsate:2012ky}, where the same model star appeared with different
a equation of state (EOS) due to the modification of gravity.
(This particular modification seems to be undetectable by the discussed
universal relations, too \cite{Sham:Lin:Leung:2013}.)
The parameter estimation (spins) through gravitational wave
observations would likely be spoiled considerably \cite{Yagi:Yunes:2013:1} for similar reasons.

A very important observation, which partly breaks this degeneracy, was recently made in
Refs.\ \cite{Yagi:Yunes:2013:1, Yagi:Yunes:2013:2}.
Relations among various measurable quantities depending on the inner
structure of NSs were found to be universal
among many proposed NS models. This includes dimensionless quantities related to the moment of
inertia, spin-induced quadrupole \cite{Laarakkers:Poisson:1999}, and tidal-induced
quadrupole \cite{Hinderer:2007, Damour:Nagar:2009:4, Binnington:Poisson:2009} of the NS. A limitation of the work in Refs.\ \cite{Yagi:Yunes:2013:1, Yagi:Yunes:2013:2}
is the use of the slow rotation approximation, in which
these quantities do not depend on the magnitude of rotation,
whereas this does not hold true for rapid rotation.
However, the slow rotation approximation should be sufficiently accurate
for many near-future measurements.

One purpose of the present work is to study an extension of the
relation between the moment of inertia ($\I$) and the spin-induced quadrupole ($\Q$)
beyond the slow rotation approximation.
A first study was done in Ref.\ \cite{Doneva:etal:2013},
where the $\I$-$\Q$ relation was considered as a
function of the observationally important (but dimensionful)
frequency of rotation. Reference \cite{Doneva:etal:2013} explores at which frequency
the $\I$-$\Q$ relation of Refs.\ \cite{Yagi:Yunes:2013:1, Yagi:Yunes:2013:2} is modified.
The universality among different NS models seems to be lost for rapid rotation.
Contrary to this expectation, we find that when the rotation
is characterized by a dimensionless parameter, the universality still holds remarkably well.
We consider three different parameters, one based on the angular momentum and two based
on the rotation frequency, where one is made dimensionless by the mass and the
other by the radius. As a consequence, the universal relation containing the
latter can further be used to infer the radius of the NS, making it an effective
tool in analyzing astronomical data.
We make this explicit by formulating a universal fit of the radius in terms of the
mass, pulsar frequency, and spin.
On a more theoretical level, we show that even for certain polytropic
EOS the universality is still present for rapid rotation.
This calls for a fundamental explanation using analytic arguments.
 
The universal relations are important since they connect several crucial astrophysical parameters. For example, the frequency and the mass are observable in binary pulsar systems using pulsar timing, and the spin  (or moment of inertia) might become measurable in the near future; see, e.g., Ref.\ \cite{Kramer_2009}. The latter two also leave an imprint in the emitted gravitational waves and can be inferred from future detection; see, e.g., Ref.\ \cite{Nielsen:2012}. The radius of a NS is measured using photospheric radius expansion and transiently accreting NSs in quiescence; see Ref.\ \cite{Oezel_2013} for a review. Both methods yield accuracies of about $10\%$ employing different models for the data analysis; see Ref.\ \cite{Steiner_2013}.


Besides the work in Ref.\ \cite{Doneva:etal:2013}, the (in)validity
of the universal relations was investigated in other regimes, too.
Nonlinear and dynamical aspects of the tidal-induced quadrupole
were included in Ref.\ \cite{Maselli:Cardoso:Ferrari:Gualtieri:Pani:2013}, where it was found
that the universal relations still hold true. In Ref.\ \cite{Haskell:2013vha}, the effect of
magnetic fields was included and it was shown that the universal relations
are broken for NSs with a large rotation period ($\gtrsim 10\,\text{s}$) and strong
magnetic fields ($\gtrsim 10^{12}\,\text{G}$) in a twisted-torus field configuration.
As an example, Ref.\ \cite{Haskell:2013vha} speculates that breaking of the universality
might just start to play a role for the slower rotating NS in the double pulsar at the time of
merger. However, one should also admit that a rotation period of $\gtrsim 10\,\text{s}$
implies that the dimensionful quadrupole will be very small and likely irrelevant
for most observations.
Various other universal properties of NSs were discussed
before \cite{Lattimer:2000nx, tsui:2004, Urbanec:Miller:Stuchlik:2013}
and after \cite{Baubock:Berti:Psaltis:Ozel:2013, Pappas:2013naa, Yagi:2013}
the discovery of Yagi and Yunes
\cite{Yagi:Yunes:2013:1, Yagi:Yunes:2013:2}.

\section{Rotating NS}
The spacetime of a rotating NS can be written in the following form (in units
where $G=c=1$):
\bea
ds^2 &=& -e^{2\nu} dt^2+ r^2 \sin^2 \theta B^2 e^{-2\nu} \left( d\phi - \omega
dt \right)^2 \nonumber\\
&&+ e^{2(\xi-\nu)}\left( dr^2 + r^2 d\theta^2 \right),
\eea
where $\nu$, $B$, $\omega$ and $\xi$ depend only on $r$ and $\theta$. 

The matter field describing the interior of the NS is modeled by a perfect
fluid of the form
\be
T^{\mu\nu} = (\rho + P)u^{\mu}u^{\nu} + P g^{\mu\nu},
\ee
where $\rho$ is the energy density, $P$ is the pressure, and $u^{\mu}$ is the
$4$-velocity. The model is specified once a particular EOS is
given as described in the next section.

To solve the field equations for the rotating NS, we use the \texttt{RNS} code \cite{Stergioulas:Morsink:1999, Stergioulas:1994ea} (see also
Ref.\ \cite{Komatsu:1989zz} for details on the method and equations) including our own
modification implementing the multipole extraction described in
Ref.\ \cite{Pappas:Apostolatos:2012}.  Note that the \texttt{RNS} code assumes rigid rotation; see Refs.\ \cite{Stergioulas:Morsink:1999, Stergioulas:1994ea}.

The metric functions, which allow us to define the quadrupole, have the
following asymptotic decay \cite{Butterworth:1976}
\bea
\nu &=& -\frac{M}{r} + \left( \frac{B_0 M}{3} + \nu_2P_2  \right)\frac{1}{r^3}
+\Order(r)^{-4} , \\
B &=& 1 + \frac{B_0}{r^2} +\Order(r)^{-4} , \,
\omega =\Omega\left(1+\frac{2I}{r^3}\right)
+\Order(r)^{-4} , \nonumber
\eea
where $M$ is the mass of the star, $P_2$ is a Legendre polynomial, and $\nu_2$,
$B_0$, $I$ are real constants, $I = J/\Omega$ is the moment of inertia, $J$ being
the angular momentum, and $\Omega \equiv 2 \pi f$ is the angular frequency measured by a distant observer
(pulsar frequency).
The quadrupole moment $Q$ \cite{Pappas:Apostolatos:2012} is then given by
\be
Q = -\nu_2 -\frac{4}{3}\left( \frac{B_0}{M^2} +\frac{1}{4} \right).
\label{defj}
\ee
We plan a multipole extraction using source integrals
as envisaged in Ref.\ \cite{Gurlebeck:2012} in the future.

In order to investigate universal relations, we introduce
dimensionless quantities as
\begin{equation}
\begin{split}\label{defIhat}
&a = \frac{J}{M^2} , \quad
\I = \frac{I}{M^3} , \quad
\Q = - \frac{Q}{M^3 a^2} , \\
&\R = \frac{2 R}{M} , \quad
\fM = 200 M f , \quad
\f = 20 R f ,
\end{split}
\end{equation}
where $R$ is the equatorial radius of the NS.
The dimensionless frequencies are such that $f=1$~kHz corresponds
to $\fM \approx 1 \approx \f$ for $M=M_{\astrosun}$ or $R=15$~km, respectively.
Similarly, it holds $\R \approx R / \text{km}$ for $M=1.4M_{\astrosun}$.


\section{EOS}
The interior structure of a NS is modeled by an EOS,
giving relations between thermodynamical quantities, such as the energy
density $\rho$ and the pressure $P$. Our lack of knowledge of high-density
nuclear matter is manifested through a series of candidate EOS.
The dependence of the interior structure of the NS on the
respective EOS 
in turn affects the exterior properties.
For example, different EOSs predict different relations between the mass and the radius,
and one might expect the same for the moment of inertia and the quadrupole.
Note that recently the radius was measured sufficiently accurate with photospheric radius expansion and quiescent low-mass x-ray binaries to
yield important constraints on the EOS in a wide range of NS masses \cite{Steiner_2013}.





Our selection of realistic (tabulated) EOS is APR \cite{akmal:1998},
AU (called AV14+UVII in Ref.\ \cite{WFF}),
FPS \cite{fps1},
BSK20 \cite{Goriely:Chamel:Pearson:2010, Pearson:2011, Pearson:Chamel:Goriely:Ducoin:2012}, and
SLy \cite{douchin:2001}.
We also include polytropic EOS $P = K \rho^{1+1/n}$
with polytropic indices $n = 0.5$ and $n = 0.6$, which we denote
by p1 and p2, respectively.
Here, $K$ introduces an irrelevant scale, which cancels in the dimensionless quantities.



\section{\texorpdfstring{Universal $\I$-$\Q$ relation for arbitrary rotation}{Universal I-Q relation for arbitrary rotation}}
\begin{figure}
\centering
\includegraphics{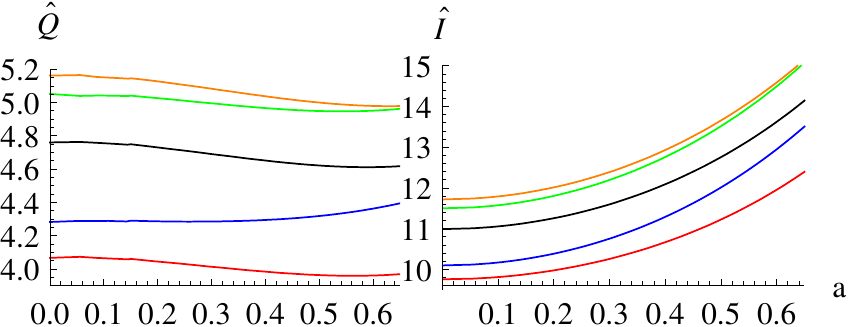}
\caption{Dependence of $\I$ and $\Q$ on the spin parameter $a$ for a
$1.4M_{\astrosun}$ NS and different EOS, from top to bottom:
BSK20, SLy, APR, FPS, AU.}
\label{fig:Ia-Qa}
\end{figure}

In the present section, we aim to study the surprising $\I(\Q)$ universality beyond the
slow rotation approximation.
The accuracy of the universality in that approximation
is better than 1\% \cite{Yagi:Yunes:2013:1, Yagi:Yunes:2013:2}.
Figure \ref{fig:Ia-Qa} indicates a deviation from the slow rotation case
within this accuracy at about $a \sim 0.1$. We will see that this is indeed the
threshold where a modification of the universal relation is required.

We switched to a low order in the angular expansion in the \texttt{RNS} code for $0.05<a<0.15$ in Fig.\ \ref{fig:Ia-Qa}, because the higher orders produce
a lot of numeric noise for small $a$. The data points at $a=0$ were obtained
in the slow rotation approximation. We checked that we can use the default
order of the angular expansion for $a>0.1$ for our desired precision goal.
This is an important aspect of our investigation, as we can smoothly connect to
the result in Refs.\ \cite{Yagi:Yunes:2013:1, Yagi:Yunes:2013:2}.
This is computationally challenging, since a large grid size is required
to stabilize the result. 

In Ref.\ \cite{Doneva:etal:2013}, a deviation from the slow-rotation result greater than $1\%$ 
showed up for frequencies between $160$ and $480 \,\text{Hz}$.
(However, the deviations become weaker as one approaches the maximum mass of the NS model.)
The natural next step is to explore if universality can be extended to this regime and beyond.
This requires a suitable dimensionless parameter characterizing rotation,
say $\alpha$, such that the relation $\I(\Q, \alpha)$ is approximately universal
among various EOSs.
Indeed, we have defined several natural candidates for such parameters
in Eq.\ (\ref{defIhat}): $a$, $\fM$, and $\f$.

\begin{figure}
\begin{center}
\includegraphics{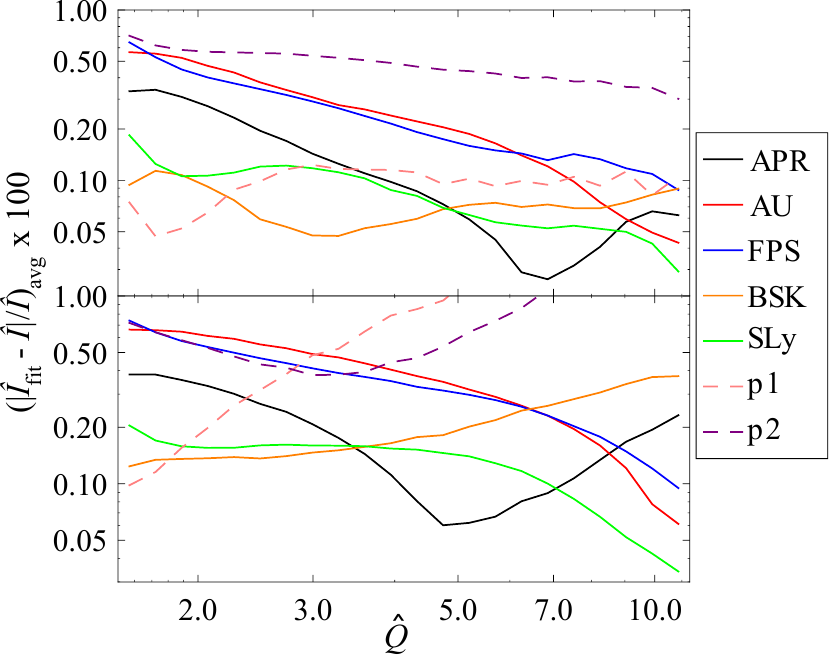}
\caption{Percentage deviation of the $\I$-$\Q$-$a$ (top) and $\I$-$\Q$-$\f$
(down) fits with respect to data points,
averaged over the $a$ (top) or $\f$ (down) direction. The deviation is almost
constant in the $a$ or $\f$ direction.}
\label{fig:aploterror}
\end{center}
\end{figure}

We extend the fit in Ref.\ \cite{Yagi:Yunes:2013:1} by a dependence on $a$ or $\f$ as
\be\label{IQfits}
\log \I \approx \sum_{i,j}  \mathcal A_{ij} a^i \log^j \! \Q
\approx \sum_{i,j}  \mathcal B_{ij} \f^i \log^j \! \Q ,
\ee
where the coefficients are given in Table \ref{tab:fits}.
We used  around 30k data points for the regime $0.1<a<0.6$, $0.2<\f<1.2$,
$1.5<\Q<15$. The deviation from these fits is maximally $\sim 1\%$ (independent of the EOS) and on average $\sim 0.3\%$. 
Figure \ref{fig:aploterror}
shows the accuracy of the fit for the selected EOS.
%
%
At the time of writing this {\paper}, we became aware of
Ref.\ \cite{Pappas:2013naa}, where a similar fit with $a$ as a parameter was given
but the discussion therein focused on other universal relations.

\begin{table}
\caption{Numerical coefficients for the fits of Eqs.\ (\ref{IQfits}) and (\ref{radfit}).\label{tab:fits}}
\begin{tabular}{c|ccccc}
	   $i =$      & 0    & 1    & 2    & 3    & 4\\
\hline
 $\mathcal A_{i0}$  & 1.35 & 0.3541 & -1.871 & 3.034 & -1.860 \\
 $\mathcal A_{i1}$  & 0.697 & -1.435 & 8.385 & -14.75 & 10.05 \\
 $\mathcal A_{i2}$  & -0.143 & 1.721 & -9.343 & 18.14 & -12.65 \\
 $\mathcal A_{i3}$  & 0.0994 & -0.8199 & 4.429 & -8.782 & 6.100 \\
 $\mathcal A_{i4}$  & -0.0124 & 0.1348 & -0.7355 & 1.460 & -1.008 \\
\hline
 $\mathcal B_{i0}$  & 1.35 & 0.1570 & -0.3244 & 0.09399 & 0.02863 \\
 $\mathcal B_{i1}$  & 0.697 & -0.6386 & 1.509 & -0.6932 & 0.05381 \\
 $\mathcal B_{i2}$  & -0.143 & 0.7711 & -1.636 & 0.8434 & -0.1210 \\
 $\mathcal B_{i3}$  & 0.0994 & -0.3594 & 0.7482 & -0.3079 & 0.06019 \\
 $\mathcal B_{i4}$  & -0.0124 & 0.05788 & -0.1140 & 0.05262 & -0.03466 \\
\hline
 $\mathcal C_{i0}$  & 3.081 & -0.1108 & 0.3402 & & \\
 $\mathcal C_{i1}$  & 0.6266 & -0.01873 & 0.08047 & & \\
 $\mathcal C_{i2}$  & -0.009608 & 0.01382 & -0.02374 & &
\end{tabular}
\end{table}

Note that the polytropes were not included in the data for the fit
but are contained in Fig.\ \ref{fig:aploterror}.
It is well known that one can approximate the EOS of a NS with polytropes in the range $n\sim 0.5 \dots 1$; see, e.g., Ref.\ \cite{Lattimer:2000nx}.
Typically, the tabulated EOSs have an $n$ value closer to $0.5$ in the core, and then it increases up to $1.0$.
Keeping this in mind, we found that for $n \lesssim 0.6$ the polytrope is in perfect agreement with our fits
(Fig.\ \ref{fig:aploterror}), whereas
the $n=1$ polytrope has greater deviation. This was observed in
the slow rotation approximation in Ref.\ \cite{Yagi:Yunes:2013:2}, too.
Therefore, polytropes can be an ideal toy model to investigate
the underlying mechanism of the universality analytically
(see, e.g., Ref.\ \cite{Stein:Yagi:Yunes:2013}).

Subsequently, we discuss three choices of the dimensionless parameters
and their implications for the universality of the $\I$-$\Q$ relation.

\begin{figure}
\begin{center}
\includegraphics{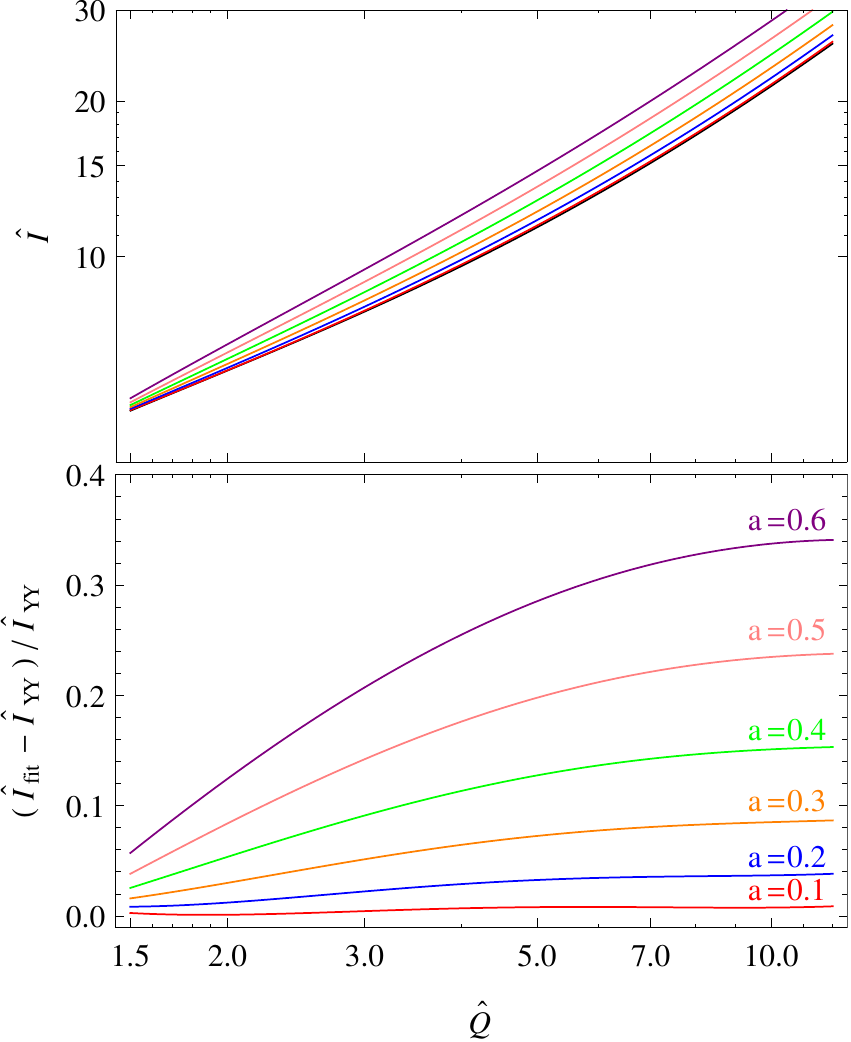}
\caption{Dependence of the $\I$-$\Q$ relation on the spin parameter $a$ (upper
plot)
and the relative deviation from the slow rotation result $I_{\text{YY}}$ (lower
plot).
The curve for $a=0.1$ is almost identical to $I_{\text{YY}}$.}
\label{fig:aplot}
\end{center}
\end{figure}

\emph{1) $a = J/M^2$ as dimensionless parameter.}
This parameter is the natural choice and works best for the proposed universality.
For fixed $a$, the
$\I$-$\Q$ relation depends on the EOS within less than 1\%.
However, it depends on $a$; see Fig.\ \ref{fig:aplot}. A simultaneous
measurement of $\I$, $\Q$, and $a$ must be consistent with these curves if
general relativity holds. This can be used to test strong-field gravity
independent of assumptions on the EOS.

\begin{figure}
\begin{center}
\includegraphics{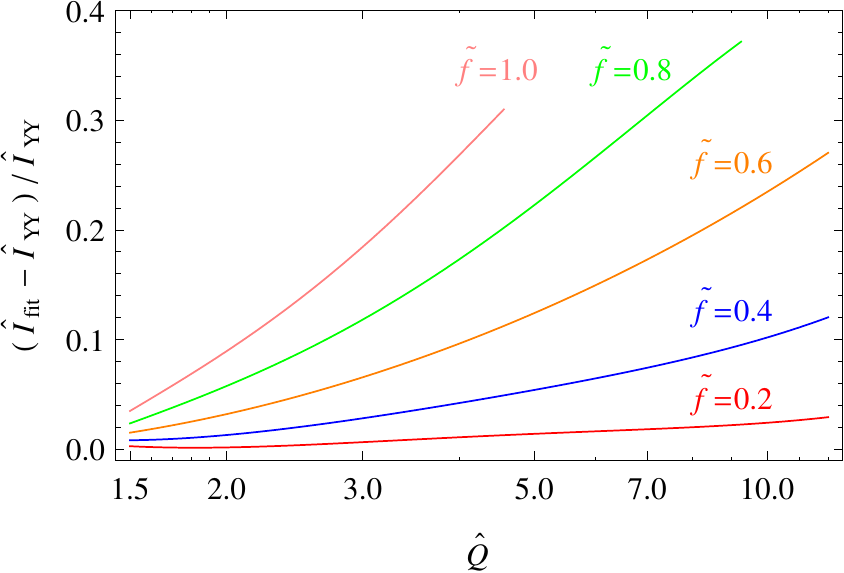}
\caption{EOS-independent deviation of the $\I$-$\Q$ relation from the slow
rotation
limit $I_{\text{YY}}$ for fixed $\f$.}
\label{fig:Rfplot}
\end{center}
\end{figure}

\emph{2) $\f \propto R f$ as dimensionless parameter.}
Again, universality is found for constant $\f$; see Fig.\ \ref{fig:Rfplot}.
In principle, this can be used to (indirectly) constrain the radius $R$ of the
NS once
the dimensionless $\I$, $\Q$, and the dimensionful $f$ (pulsar frequency)
are known.
This is an important prospect of this {\paper}, since a direct measurement of $R$ is difficult. 
But the mass-radius relation contains invaluable information about the EOS.
Thus, although the $\I$-$\Q$ relation is universal among EOS, it can still be
useful to constrain them.

\emph{3) $\fM \propto M f$ as dimensionless parameter.}
This is not independent from the parameter choice 1).
From Eq.\ \eqref{defIhat} along with the definition of $I$, it can
easily be seen that $\I \propto a/\fM$.
Interestingly, the lines of constant $\fM$ look quite different from Fig.\ \ref{fig:aplot}.
Instead, they qualitatively resemble Fig.\ 1 in Ref.\ \cite{Doneva:etal:2013}, but
display good universality now.

\section{Combination of relations}
\begin{figure}
\begin{center}
\includegraphics{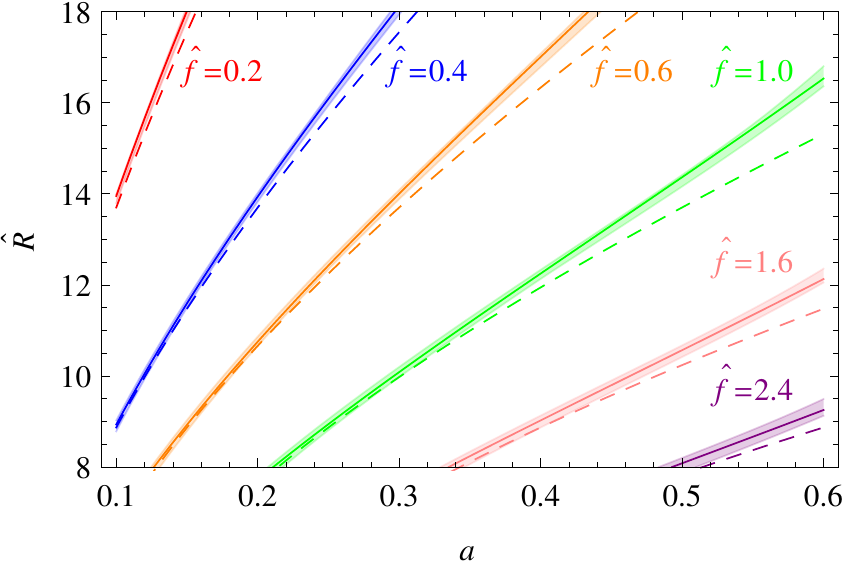}
\caption{Fit of Eq.\ (\ref{radfit}) including deviations from our data (shaded area)
and the slow rotation fit from Ref.\ \cite{Baubock:Berti:Psaltis:Ozel:2013}
(dashed lines).\label{fig:Rafplot}}
\end{center}
\end{figure}
The aforementioned relations can of course be combined. In particular, 
one can solve relation 1) for
$\Q$ (i.e., use it to ``measure'' $\Q$), eliminate it from relation 2), and obtain
an $\I$-$\f$-$\fM$ relation. This is useful, as $Q$ is most difficult to measure.
Next, identities among the dimensionless quantities ($a \propto \I \fM$, $\f \propto \R \fM$)
allow a reformulation as an $a$-$\R$-$\fM$ relation, which we fit as
\be\label{radfit}
\log \R \approx \sum_{i,j}  \mathcal C_{ij} a^i \log^j \! \frac{a}{\fM}.
\ee
The result is depicted in Fig.\ \ref{fig:Rafplot}.
The maximal deviation in Fig.\ \ref{fig:Rafplot} is about 2\% for our selection of EOS.
However, Eq.\ \eqref{radfit} does not fit very well to the data generated using
the polytropes discussed above. Hence,
one must expect an increasing deviation if further EOSs are
included in the future.
Still, this relation should put tight constraints on $R$ if
$a$, $f$, and $M$ are known.
We find that a reformulation of the slow rotation fit given in Eq.\ (15) of
Ref.\ \cite{Baubock:Berti:Psaltis:Ozel:2013} is in good agreement with our
Eq.\ (\ref{radfit}) --- surprisingly even in the rapid rotation regime.
Since our fit uses the observable mass
and equatorial radius, this extrapolation of the slow-rotation case is indeed astonishing.
Other combinations of the relations can be studied; e.g., one can try to
formulate a mass measurement (given a radius) or eliminate the frequency
(for pure GW observations).

The universal relation in Eq.\ \eqref{radfit} can either be used to measure or constrain
the radius or to improve the accuracy of the radius measurement. If future x-ray
observatories increase the accuracy of the radius measurement sufficiently, then
this relation can be used to test fundamental physics. This is what makes
universal relations, its combinations, and reformulations so powerful: One can
use them to infer unobservable properties or to test gravity if all
quantities entering the relations are observable.
The diversity of upcoming instruments (see Introduction)
makes this even more interesting.

\acknowledgments
We acknowledge useful discussions with K.\ Yagi, N.\ Stergioulas, G.\ Pappas, P.\
Pani and our colleagues of NESTAR.
We are very grateful to N.\ Chamel for supplying EOS tables and
for useful discussions.
This work was supported by DFG (Germany) through projects STE 2017/1-1 and STE 2017/2-1,
FCT (Portugal) through projects PTDC/CTEAST/098034/2008, PTDC/FIS/098032/2008,
SFRH/BI/52132/2013, and PCOFUND-GA-2009-246542 (co-funded by Marie Curie Actions),
and CERN through project CERN/FP/123593/2011.
The authors thankfully acknowledge the computer resources, technical
expertise, and assistance provided by CENTRA/IST. Computations were
performed at the cluster ``Baltasar-Sete-S{\'o}is'' and supported by the
ERC Starting Grant No.\ DyBHo-256667.
We gratefully acknowledge support from the DFG within the Research
Training Group 1620 ``Models of Gravity.''

\ifnotprd
\bibliographystyle{utphys}
\fi

\ifarxiv
\providecommand{\href}[2]{#2}\begingroup\raggedright\endgroup

\else
\bibliography{iloveq}
\fi

\end{document}